\documentclass[aps,prl,citeautoscript,twocolumn,floatfix,showpacs]{revtex4} 
\usepackage{graphicx}
\usepackage{amsmath}
\usepackage{subfigure}
\usepackage{dcolumn}% Align table columns on decimal point
\usepackage{bm}% bold math

\begin{document}

\title{Fixed-node errors in  
quantum Monte Carlo: interplay of electron density and node nonlinearities}
% testing comment
\author{Kevin M. Rasch, Shuming Hu, Lubos Mitas}
\affiliation{
Center for High Performance Simulation and Department of Physics,
North Carolina State University, Raleigh, NC 27695}

\date{\today}

\begin{abstract}
We elucidate 
	the origin 
		of large differences (two-fold or more)
			in the fixed-node errors 
				between the first- vs second-row systems
				for single-configuration trial wave functions
in quantum Monte Carlo calculations.
This significant difference
	in the fixed-node biases 
is studied 
	across a set of atoms, molecules, and also Si, C solid crystals.
The analysis is done 
	over valence isoelectronic systems 
		that share similar correlation energies, bond patterns, geometries, ground states, and symmetries. 
We show that 
	the key features which affect the fixed-node errors
	are the differences 
		in electron density and the degree of node nonlinearity.
The findings reveal how the accuracy 
of the quantum Monte Carlo varies 
		across a variety of systems, 
	provide new perspectives
		on the origins of the fixed-node biases 
		in electronic structure calculations of molecular and condensed systems, 
	and carry implications
		for pseudopotential constructions for heavy elements.  
	
\end{abstract}

\maketitle

Quantum Monte Carlo (QMC) approaches have proven to be remarkably successful in
studies of many-body quantum systems.
For electronic structure calculations, 
QMC offers an important alternative~\cite{qmcrev,qmcrpp} 
to approaches based on either density functional theory 
(DFT) or basis set correlated wave function methods.
The fundamental inefficiency of fermion signs in QMC can be 
overcome by the fixed-node approximation in which
the node (zero locus) of a desired stationary eigenstate is 
constrained to be identical to the
node of the best available trial wave function.
It is very encouraging that commonly used 
 Slater-Jastrow trial wave functions 
enable us to recover about 90-95\% of the correlation
energy almost universally, in atoms, molecules, solids, quantum liquids, etc.
Energy differences produced 
by this ``QMC standard model,'' 
such as bindings/cohesions,
reaction barriers,
and excitations, for example, agree with experiments within a few percent for many systems
and elements across the periodic table~\cite{qmcrev,qmcrpp}.

Despite these achievements, the fixed-node error remains the key limiting factor in  
the quest for higher systematic accuracy 
and in studies of quantum phenomena at finer energy scales.
Several methods for improving nodal hypersurfaces have been proposed 
(see, e.g., ideas in recent Ref.~\cite{shdmc}), however,
it is fair to say that such effort is almost invariably very demanding and 
often not fundamentally different from slowly converging 
wave function expansions in excitations. Our understanding of the nodal 
hypersurfaces is still very rudimentary and other than insights
into topologies of nodal cells \cite{ceperley,dario,mitas}, 
many aspects of these remain unknown. Therefore,
identifying the source of the nodal inaccuracies in the most commonly
used variational wave functions is perhaps the most important outstanding 
question.

Recently, high-accuracy QMC calculations 
showed a striking difference between 
the fixed-node errors of  Si$_2$ vs C$_2$ dimers~\cite{qmcc2}.
Interestingly, this turned out to be just a special case of a broader and more systematic trend.
As we show below,  
for commonly used single-configuration Slater-Jastrow trial functions,
the fixed-node errors of the first- vs second-row atom systems systematically differ by 
a significant amount, i.e., by a  factor of two or more.
This difference persists in atoms, molecules, and solids 
despite the fact that the compared systems 
have identical ground states and symmetries, 
similar bonding patterns, geometric structures, valence correlation energies,
and even qualitatively similar excitation spectra.
Our analysis of this effect is built on recent findings 
that in simple atomic systems the nodal errors grow in proportion
to the electronic density~\cite{qmcdens}
(and similar result was found for homogeneous 
electron gas~\cite{qmcala} with the error 
being proportional to $\ln r_s$).  
Ultimately, we reveal that besides the electron density, 
the fixed-node biases are strongly
affected by the node nonlinearities. These nonlinearities originate from 
the different degree of localization of 
occupied states in different symmetry channels, and they are further affected 
by bonding patterns and bond multiplicities.   
Our findings have rather wide implications for systems composed from
 main group elements,
 3$d$ transition metals, and beyond.
Our analysis has potential implications also
for construction of pseudopotentials which can alleviate some of these effects in heavier
elements.

{\em QMC calculations.} Total energies of a selected 
set of first- and second-row 
systems have been calculated by fixed-node diffusion Monte Carlo
(FNDMC) with single-configuration Slater-Jastrow trial functions. 
The only exception was the C atom, where we 
included two configurations due to the known and sizable 
 $s^2 \to p^2$ near-degeneracy, which does not affect 
the rest of compared systems (e.g., in the Si
atom it is below $\approx$ 1 mHa). 

For generating one-particle orbitals 
we have employed Hartree-Fock (HF) 
and several DFT functionals, including 
hybrids; and for each system we have chosen the orbital set
 with the lowest fixed-node energy. For all elements except hydrogen,
we have used the energy consistent pseudopotentials (PPs)~\cite{ecpbfd} and 
basis sets of  the cc-pV5Z quality. Our total energies are similar to or 
marginally higher than (1mHa or less) 
the recent high-accuracy calculations of the G2 benchmark set of molecules that
additionaly included
explicit QMC orbital reoptimizations \cite{g2new}.
QMC calculations of the C and Si solids in the diamond structure 
were done with 8- and 64-atom supercells with 
twisted averaging, extrapolation to the thermodynamic
limit using the $S({\bf k})$ correction~\cite{chiesa}, 
and for the Si crystal we verified that 216-atom 
supercell result was consistent with the extrapolation.

{\em Estimation of the exact total energies. } 
For an independent estimation of the exact total valence energies with
the target accuracy of 1mHa/atom (or better) in atoms and molecules, we have 
carried out CCSD(T) calculations with ccpVnZ basis sets with $2\leq n\leq 5$
 and extrapolated to $n\to \infty$. The extrapolation formula 
was based on the sum of two contributions: exponential for the  
HF energy component and sum of power terms $c_1n^{-3}+c_2n^{-5}$
for the correlation~\cite{feller}. This formula  provided very good
agreement with ours and also previously reported high-accuracy, multi-reference FNDMC
calculations for selected systems such as N, C  atoms and N$_2$ dimer~\cite{qmcc2,qmcmiso}. 
We found the results to agree within $\approx$ 0.6 mHa. 
Solely for the C and Si solids, we estimated the exact total energies 
with empirical inputs, using the experimental 
cohesive energies corrected for the zero-point motion.

\begin{figure}
\centering
\includegraphics[width=0.92\columnwidth,clip,angle=0]{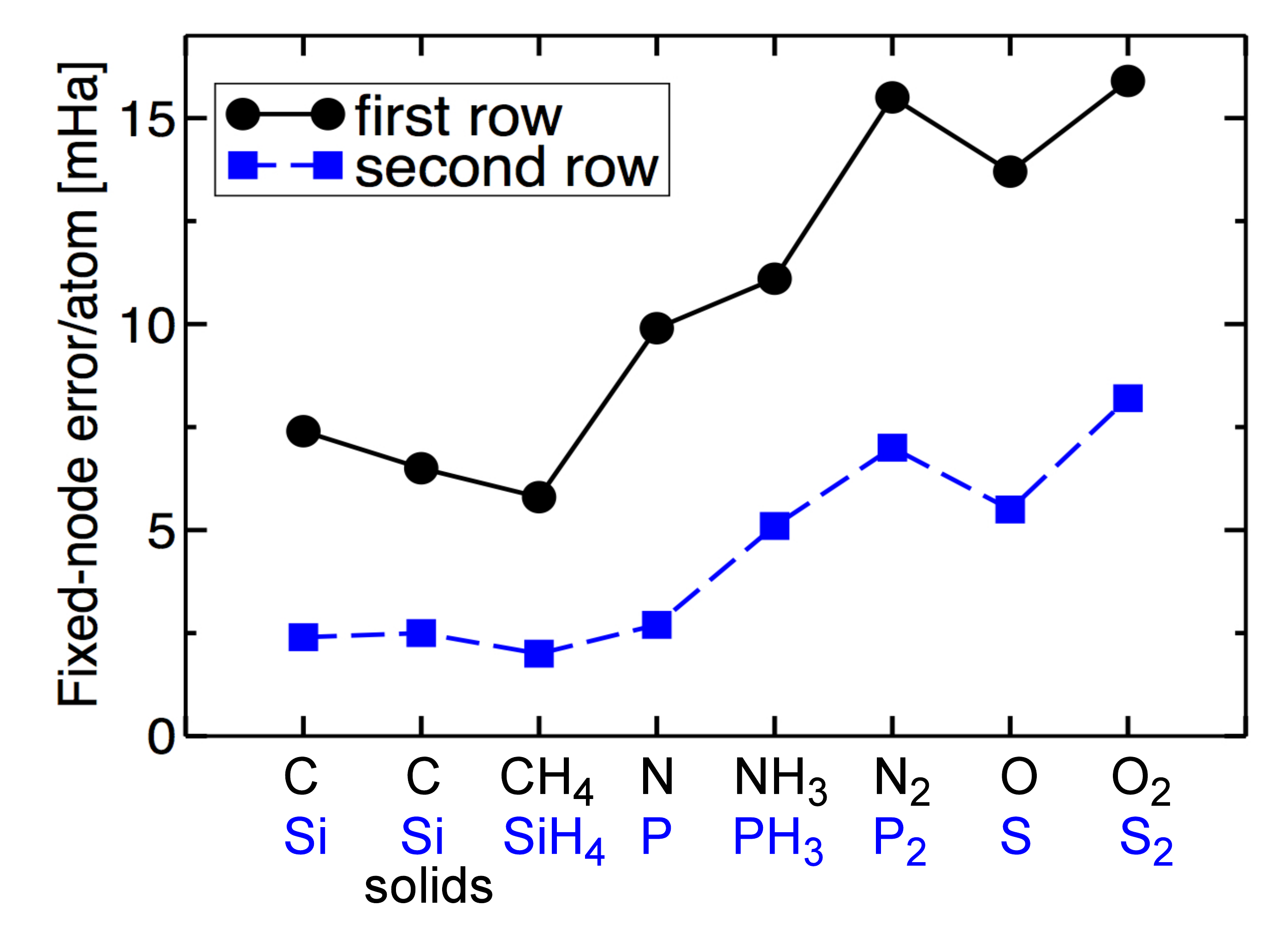}
\caption{Fixed-node errors of the single-configuration trial wave functions
for a set of first- and second-row
atoms, dimers, hydrides and the diamond structure solids of C and Si.
The values are normalized per non-hydrogen atom. 
%The triangle for the C atom
%corresponds to the single-configuration trial function, see text for details.
}
\label{fig:fnerror1}
\end{figure}

\begin{figure}
\centering
\includegraphics[width=0.77\columnwidth,clip,angle=0]{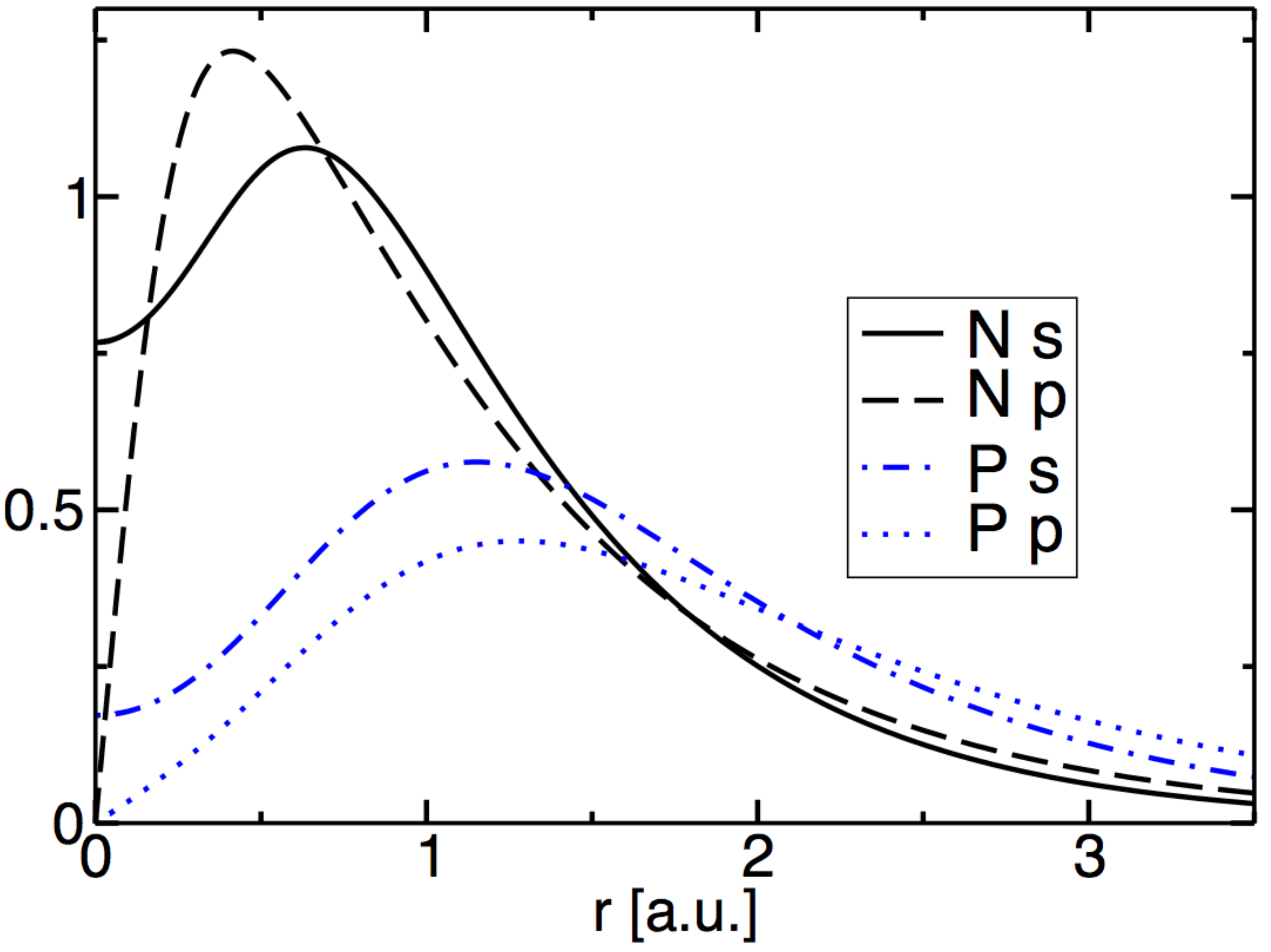}
\caption{Radial parts of $s (\ell=0) $ and $p (\ell=1)$ valence 
PP orbitals plotted as $r^{\ell} 
\rho_{\ell}(r)$ for the N and P atoms.
}
\label{fig:csiorbs}
\end{figure}

{\em Fixed-node errors.} 
The resulting fixed-node errors are summarized in Fig.~\ref{fig:fnerror1}. 
The chosen first- and second-row systems 
 are directly comparable,
 i.e., they are valence isoelectronic, share very similar 
geometries, bond patterns and have the same ground 
states and symmetries. Except for the atoms, 
we mostly present closed-shell states in order to simplify the analysis 
although the effect persists in open-shell states as well (e.g., C$_2$ vs Si$_2$
shows such a difference as well and is only further enhanced
by the multi-reference C$_2$ ground state~\cite{qmcc2}).
Our first key observation is that
all the fixed-node errors for the first row are systematically and significantly higher than for the second row.
Remarkably, this is true not only for atoms and molecules but also for the Si and C crystals.

The results in Fig.~\ref{fig:fnerror1} raise a natural question: what is the reason for this trend?
First, we consider the fact
%that valence total energies                          
%and corresponding
that the valence correlation energies are larger in the first row.
They differ only moderately,
by about 20\% or less. For example, the correlation energy of the N atom is 
$\approx134$~mHa while for the P atom it is $\approx120$~mHa, i.e., about 10\% difference. 
However, this does not explain why the fixed-node error for P is only $\approx2.5$ mHa 
($\approx$ 2\% $E_{corr}$), while for N it is
 $\approx10$ mHa ($\approx8$\% of $E_{corr}$), i.e., 
the difference is four times larger both in absolute and 
relative terms. 
The multi-reference character of the states is not an issue either
since most of the studied cases exhibit large gaps and 
the wave functions are nominally of the single-reference type.
Also note that excitations in P$_2$ lie lower than in N$_2$,  
but the error is much larger
in the latter case. That is counterintuitive since one would expect
more pronounced effects and mixing due to the smaller gap of P$_2$.
%(ie, how high is the excitation spectrum does not necessarily determine the nodal errors).
Another suggestion would be to look at pseudopotentials as the source of the 
differences.  As we will see below, pseudopotentials can and do
influence the results, but they are {\em not} the root cause. 

Let us then focus on the key qualitative difference 
between the two rows, i.e., the absence of $p$-states in cores of the first row. 
For these elements the $p$-states are much more localized in the region close to nucleus,
in particular, the orbital maximum appears at a smaller radius and is higher than for the $s$-states.
The opposite is true for main elements in the second row (and rows below),
see the examples of N and P atoms in Fig.~\ref{fig:csiorbs}.

\begin{figure}
\centering
\includegraphics[width=0.75\columnwidth,clip,angle=0]{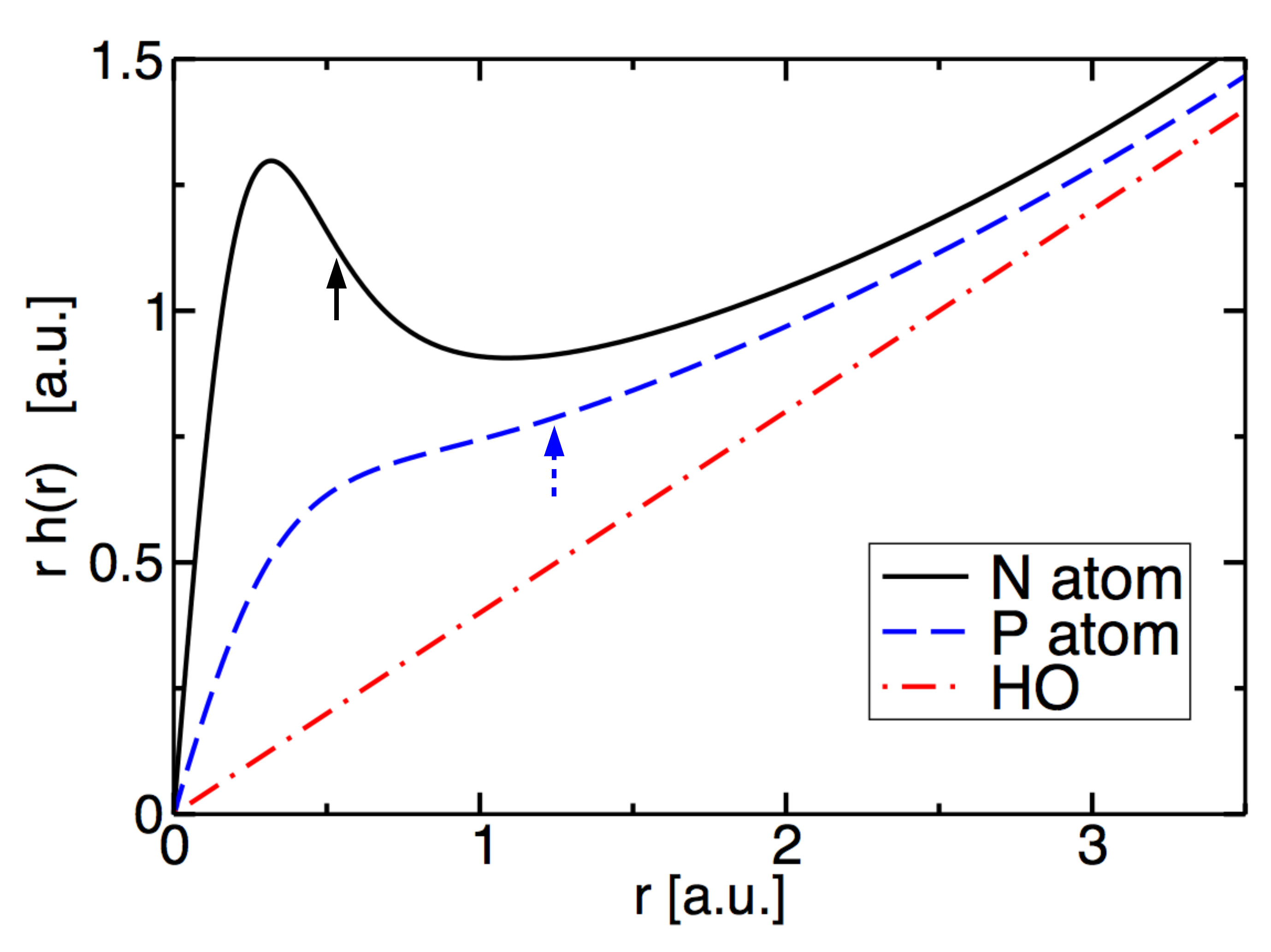}
\caption{ The functions $r h(r)=r\rho_p(r)/\rho_s(r)$ for N atom, P atom and harmonic
oscillator fermions. The small arrows indicate the core radii of N atom
(0.55 au) and P atom (1.27 au) 
defined as the outermost maximum of the one-particle electron density.
}
\label{fig:fig93}
\end{figure}

{\em The simplest nodes model and nodal analysis.}
We analyze the impact of the spatial character of $s,p$ states 
on the nodes starting from the example of
non-interacting 3D harmonic fermions. This choice is motivated by the fact
that this is perhaps the 
simplest solvable model of nodes in localized systems (the nodes turn out to be 
zeros of homogeneous antisymmetric polynomials~\cite{mitas}).  It also enables us to  
 construct states with the same symmetries as few-electron atoms.
Let us therefore write the harmonic oscillator (HO) one-particle orbitals  
as $g(r), g(r)x, g(r)y, ... $ where
$g(r)$ is a gaussian.
% (further HO details are unimportant). 
The simplest state with the fermion node
is the two-electron triplet $^3P(sp)$ and has the wave function  
$$
\Psi_{HO}(1,2) = g(r_1)g(r_2){\rm det}[1,x]= G_0 (x_2-x_1)
,
$$
where the gaussians are absorbed into the non-negative prefactor $G_0$.
%The node is determined by the antisymmetric part that
%vanishes for configurations with $x_1=x_2$. 
The node is $x_1=x_2$,
and the gradient of the antisymmetric factor is constant so that the node is flat. 
%(In fact, the node of this state is exact if the two particles interact
%by the Coulomb or other $r_{12}$-dependent potential\cite{taut}.)   
For more electrons, the wave function has the same form: 
non-negative prefactor times a homogeneous antisymmetric 
polynomial with the node being its zero locus.  

Let us now reanalyze the same $^3P(sp)$ state using 
one-particle orbitals given as 
$\rho_s(r)$, $\rho_p(r)x, ... $ from wave functions of the 
pseudized first- and second-row atoms.
By dividing out  
the nonnegative $\rho_s(r)$, we get  
$$
\Psi_{atom}(1,2) = \tilde {G}_0 \times [h(r_2)x_2 - h(r_1)x_1],
$$
i.e., the non-negative prefactor times the antisymmetric part, 
where $h(r) = \rho_p(r)/\rho_s(r)$. For simplicity, let us now assume 
that the particles 1 and 2 are confined to the $x$-axis.
It is then quite revealing to plot
the function $h(|x|)x\equiv h(r)r$. Fig.~\ref{fig:fig93} shows this function 
for the N, P
atoms and HO (rescaled by multiplicative constants for a better comparison).
Clearly, for HO it is linear since 
the corresponding $h(r)=1$. 
For the P atom the function deviates from linearity,
however, the key observation is that it remains {\em monotonous}.
This implies 
that there is only one set of configurations 
which fulfills the nodal condition
$r_1 h(r_1)=r_2 h(r_2)$, qualitatively the same as for 
 HO. The nodes of these two systems are therefore similar, 
with some deviation from flatness in the P case. 
However, this contrasts with 
$rh(r)$ for N which shows a distinctly different behavior,
with a pronounced maximum at $r \approx 0.35$ au and a minimum around
$r \approx  1.15$ au. It is important to note that the nonlinearity spans well beyond the 
core radius defined as the outermost maximum of the one-particle electronic density.
Therefore this feature has its root in the all-electron structure 
since beyond the core radius the (pseudo) orbitals and also $h(r)$ functions 
are virtually identical to their all-electron counterparts. 
The pseudopotential merely smoothes 
out their behavior inside the core.
Clearly, similar nonlinearities generated by 2$s$ and $2p$ states are present
in the second-row atoms. The key difference is
that these effects are fully contained in the core while for the first row
they {\em unavoidably extend well into the bonding region}.

\begin{figure}
\centering
\subfigure[N atom, nearly exact nodes]{%
    \includegraphics[width=0.45\linewidth,height=3.6cm]{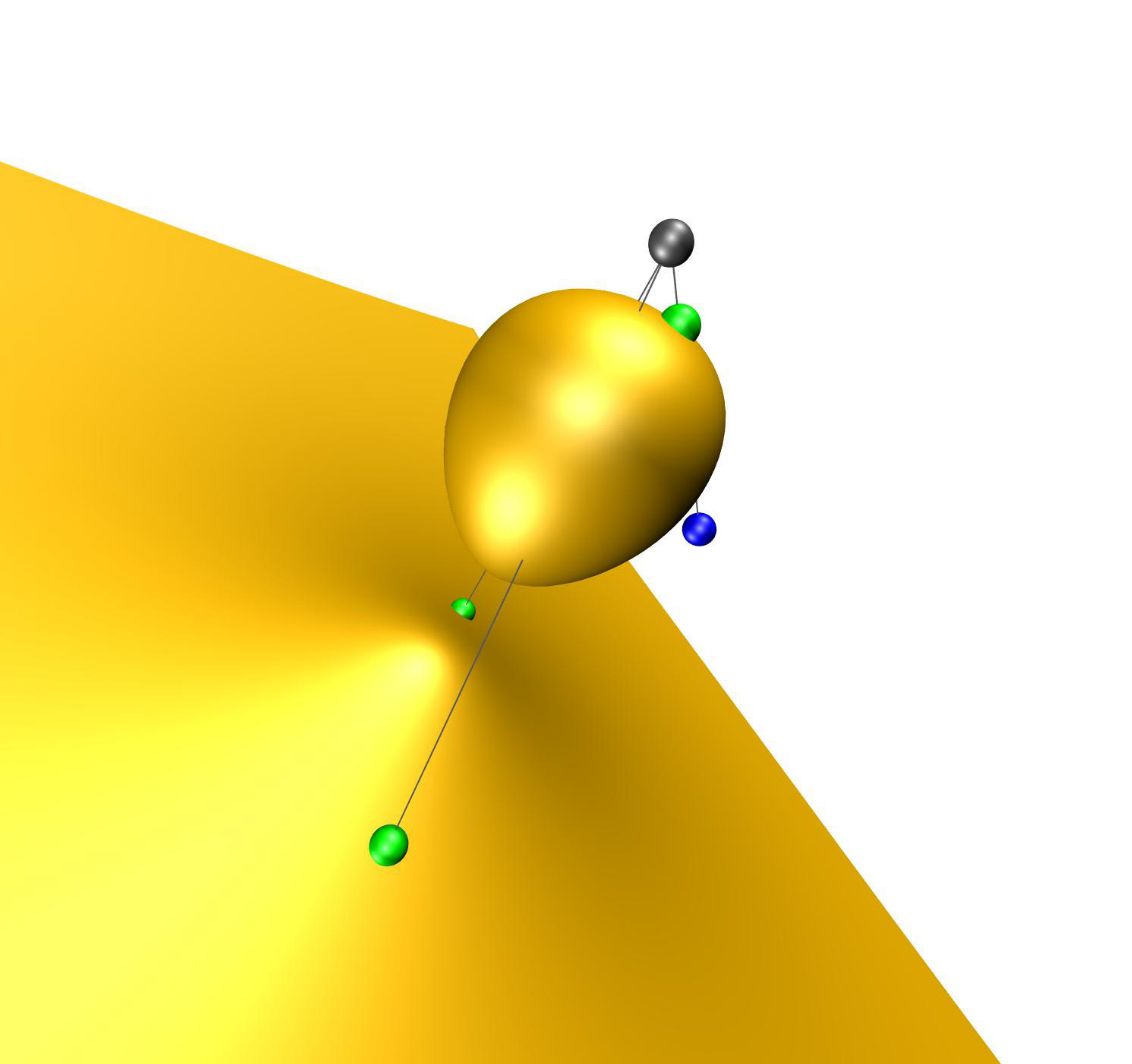}
    \label{fig:n.shdmc}
}
\quad
\subfigure[ N atom,  HF nodes]{%
    \includegraphics[width=0.45\linewidth,height=3.6cm]{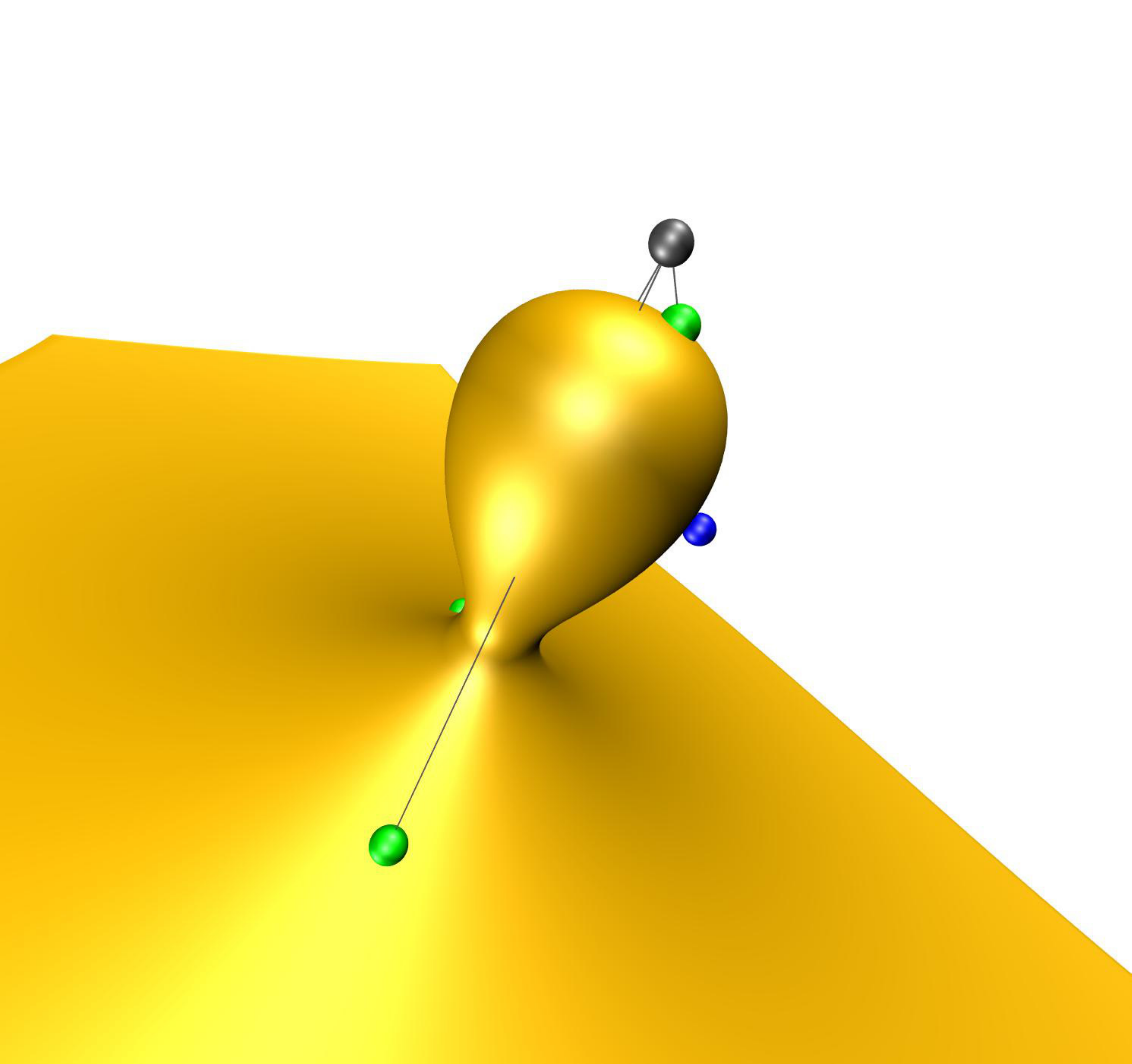}
    \label{fig:n.hf}
}
\subfigure[ P atom,  nearly exact nodes]{%
    \includegraphics[width=0.45\linewidth,height=3.6cm]{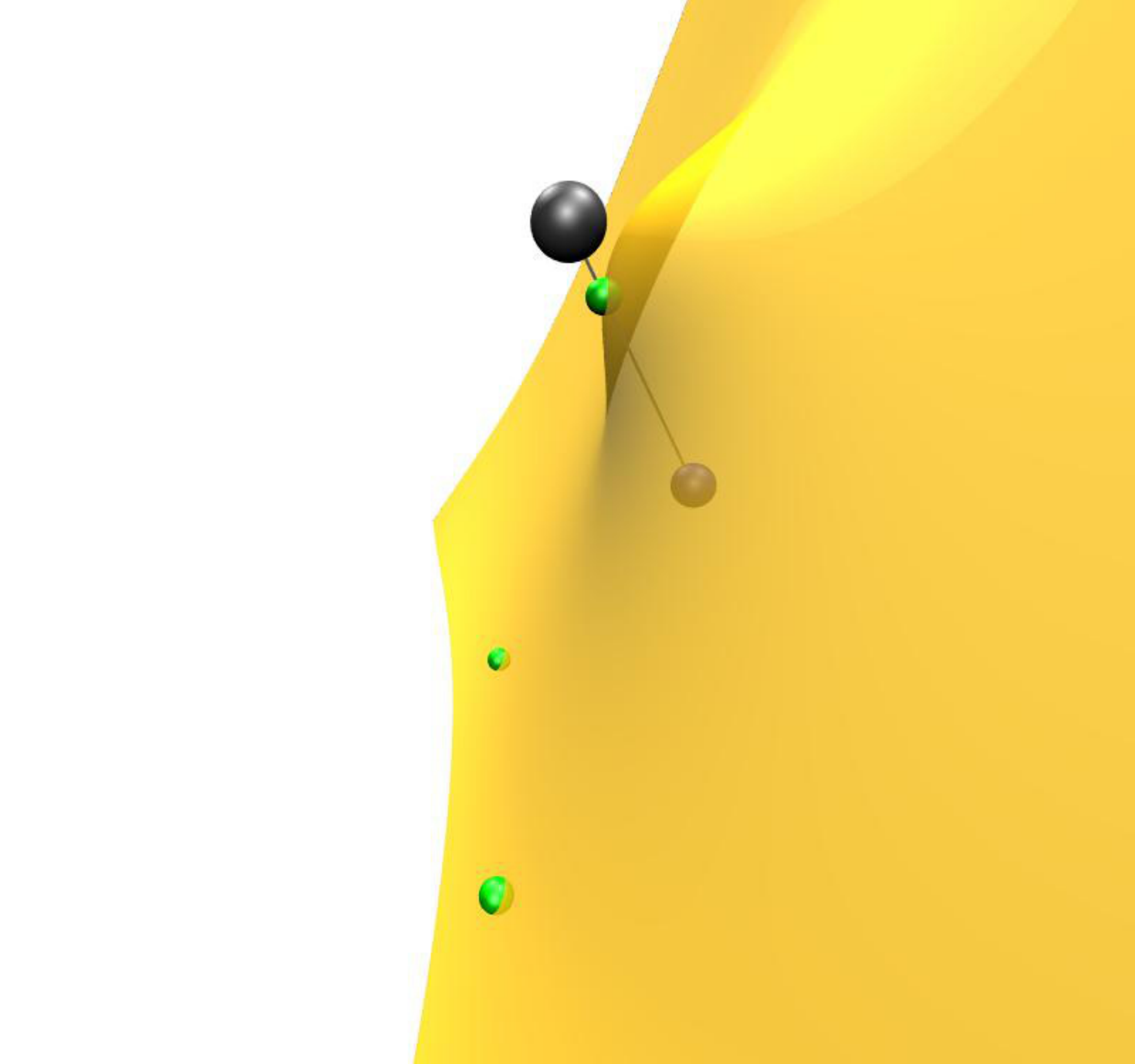}
    \label{fig:p.shdmc}
}
\quad
\subfigure[ P atom,  HF nodes]{%
    \includegraphics[width=0.45\linewidth,height=3.6cm]{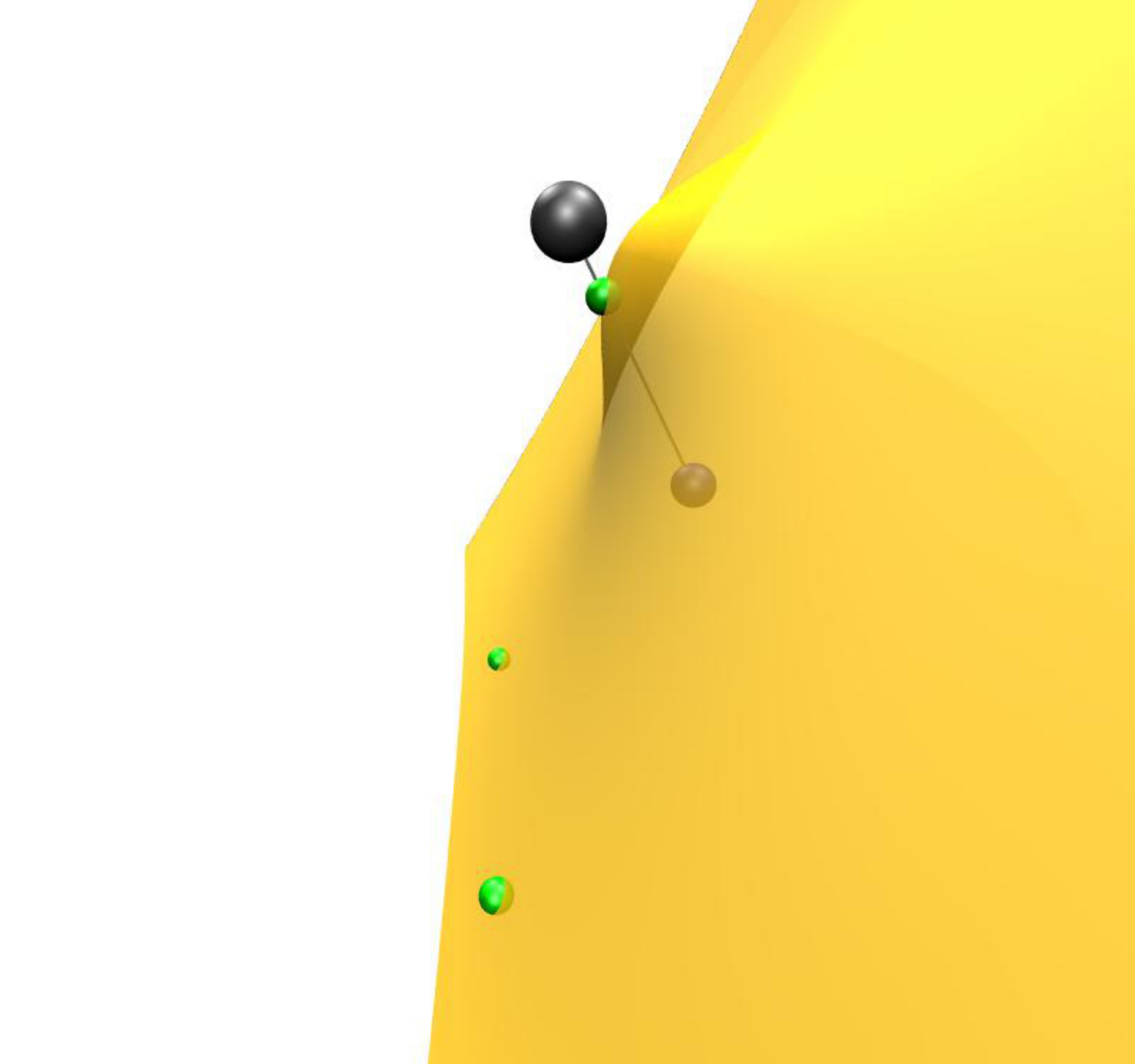}
    \label{fig:p.hf}
}
\caption[]{High curvature nodal features of the N and P atoms from 3D scans by a majority spin electron for HF and multi-reference (nearly exact, $\approx$ 1 mHa) wave functions. The larger (black) spheres are the nuclei while the smaller spheres are snapshot positions of the rest of electrons (green/blue for the majority/minority spin).  The fixed electrons are at the corresponding positions (radially adjusted according to the core radii) for both atoms 
but their nodes are dramatically different. In addition,
the N atom node is much more sensitive to interactions as seen from the difference between the
two wave functions.  
}
\label{fig:npnodes}
\end{figure}

This nonlinearity has a significant impact on the nodal structure of the wave function. 
In particular, for certain ranges of positions of electrons 1 and 2, there are multiple
sets of configurations for which the wave function vanishes.
When compared to HO, the nonlinearity of $rh(r)$ suggests that
the wave function qualitatively corresponds to
an antisymmetric polynomial of higher degree  
and the complexity of its nodal shape grows accordingly.
This analysis can be extended to  
include the rest of the electrons in full 
3D space and one finds that the first-row systems exhibit complicated nodal shapes. 
For certain configurations, distortions appear 
in the form of bulges and  
even detached bubbles, see Fig.~\ref{fig:npnodes}. 
In this Figure, we show a 3D subset of the node
found by
 3D scan of the trial function by one majority-spin electron
with the rest of the electrons fixed at snapshot positions.
These nonlinear features have high curvatures and are also sensitive 
to interactions since they are not ``held in place'' by the other electrons.  

One might argue that such nonlinearities affect only  small regions of configurations 
and since the wave function vanishes at the node the impact on
expectations should be small as well. 
However, this argument is not strictly correct, especially if we are interested
in the last few percent of the correlation energy.
 One can show that 
the wave function nonlinearities at the node are ultimately related to the 
wave function values elsewhere and therefore
have an impact on the total energy. Recently,
we have proposed to write
~\cite{nda} 
the total energy of an exact many-body eigenstate $\Psi_0$ for a general potential 
$V({\bf R})$
as follows 
$$
E_0 = \left[\int_{\partial \Omega} |\nabla\Psi_0| d{\bf S} +\int V({\bf R})| \Psi_0| d
{\bf R}\right]
/\int | \Psi_0| d{\bf R}
$$
where the first integral is over the nodal (hyper)surface $\partial\Omega$
while the other integrals
are over the particle coordinates.
Note that these are not usual expectations since 
the averages are over $|\Psi_0|, |\nabla \Psi_0|$. 
%so that the "kinetic" and "potential" components are not the usual expectation values. 
Instead, the "kinetic" component is more sensitive to the nodal features since it depends 
on the node (hyper)area and on the gradient of the wave function at the node locus. 
As we have shown elsewhere
\cite{nda}, 
 this diagnostics has enabled us to distinguish between two different nodal surfaces for 
degenerate states as well as to show, for example, equivalency of nodal surfaces for
two states with different symmetries.
Estimation of these components in QMC reveals
that the first term for the N atom is approximately 3.3 times larger
than for the P atom, reflecting thus the node differences in a
quantitative manner.  Note that atomic errors rise very 
significantly with the double
occupation of $p-$states (for example, O and S atoms), confirming that 
the impact of the electron density increase is qualitatively similar in both rows.

{\em Systems with  bonds.}
Since bonding affects both the density and  
molecular orbitals, one expects that the nodal errors will be influenced
by the degree of node nonlinearity and by the density changes
from the bond geometries. 
In particular, for Si systems with tetrahedral bonding, 
the errors are only about
2\% of the correlation energy.  This is true for {\em both} the silane and 
the solid.  In addition, we claim that this will hold also in
other Si systems with tetrahedral 
arrangements of single bonds such as Si$_2$H$_6$, etc.  Single bonds
arrangements
and low overall density therefore appears to decrease 
the corresponding fixed-node errors. 
Moreover, we expect this to be true
basically for all such IV-group element systems below the first row!
On the other hand, the errors grow with  
shortening of the single bonds in NH$_3$, PH$_3$, and they are
 the largest in systems
with multiple bonds and high densities such as N$_2$. It is instructive to
roughly estimate how much of the molecular errors originate in atoms 
and how much result from multiple bonds and associated effects.
% Note that even if we subtract the 
%errors which originate in atoms and propagate to bonded systems, the 
%resulting ``net'' bias from the multiple bonding is significant. 
Assuming that the errors are approximately additive, the effect of multiple 
bonds in N$_2$ is $\epsilon_{mol}-2\epsilon_{at} $ $\approx$ 32-2$\times$10 = 12mHa, while for
P$_2$ we get $\approx14-2\times2.5 = 9$mHa. The smaller value for 
P$_2$ reflects the longer bond length, smaller
density, and  correspondingly smaller orbital curvatures.  
It is not too difficult to accept that {\em qualitatively}
the multiple bonds not only raise the density but also increase the degree
of the node nonlinearity. In particular,
the ``stacking" of multiple bonding - and also antibonding orbitals which have 
significant densities along the bond axis -
can be approximately described as a polynomial with a higher degree than 
in isolated atoms. 
In contrast, 
note that the tetrahedral 
single bond arrangement does not increase the nodal error, for example, in methane
and diamond. As argued above, this arrangement appears to be ``optimal" and the 
existing nodal bias comes almost exclusively from the atoms, for both Si and C such systems. 

{\em Conclusions.}
We have shown that fixed-node errors in atoms
and bonded systems are influenced mainly by two
factors: the high electronic density and the
 node nonlinearity in such regions. In particular, we have found that:
%Our analysis leads to the following observations:  
a) the first-row atoms exhibit node nonlinearities that
significantly raise the fixed-node errors,
due to the high density of $p-$states close to nuclei;
b) for the main group elements in the second-row and below this
effect is almost absent; 
c) in bonded systems the fixed-node bias grows with shorter bonds 
and higher bond multiplicities. 
Perhaps the most far-reaching conclusion is that the contrast between
the fixed-node bias of the
first vs the second rows and rows beyond, holds for
 {\em all} main group ($sp$) elements across the periodic table.
 In particular,
for atoms and tetrahedral bonded systems of Si (Ge, ...) 
the fixed-node errors amount to only $\approx$ 2\% of E$_{corr}$.
The results also suggest that the 3$d$ transition 
elements exhibit similar 
effects although they are mostly confined to the core region so that systems with 
$p-d$ bonds exhibit quite accurate energy differences due 
to error cancellations~\cite{jindrafeo}. 
In addition, we envision that for heavier elements the atomic node 
nonlinearities can be alleviated by properly adjusted pseudopotential constructions. 

We believe that our analysis provides a new perspective on the origins of the fixed-node biases,
reveals why the accuracy of fixed-node QMC calculations varies for different types of systems, 
and offers hints on future opportunities to address this fundamental 
challenge.

{\em Acknowledgments.}
We would like to 
thank M. Bajdich for his efforts in the very early stages of this work.  
We acknowledge support by NSF OCI-0904794 and  
ARO W911NF-04-D-0003-0012 grants and by XSEDE
computer time allocation at TACC.


\begin{thebibliography}{99}

\bibitem{qmcrev}
W. M. C. Foulkes, L. Mitas, R. J. Needs, and G. Rajagopal,
Rev. Mod. Phys.  \textbf{73}, 33 (2001).

\bibitem{qmcrpp}
J. Kolorenc and L. Mitas, Rep. Prog. Phys.  \textbf{74}, 026502 (2011).

\bibitem{shdmc}
F. A. Reboredo, R. Q. Hood, and P. R. C. Kent, Phys. Rev. B, \textbf{79}, 195117 (2009).

\bibitem{ceperley}
D.M. Ceperley,
J. Stat. Phys.,  \textbf{63}, 1237 (1991).

\bibitem{dario}
D. Bressanini and P.J. Reynolds, Phys. Rev. Lett.  \textbf{95}, 110201 (2005).

\bibitem{mitas}
L. Mitas,
Phys. Rev. Lett. \textbf{96}, 240402 (2006).

\bibitem{qmcc2}
C. J. Umrigar, J. Toulouse, C. Filippi, S. Sorella, and R. G. Hennig,
Phys. Rev. Lett.  \textbf{98}, 110201 (2007).

\bibitem{qmcdens}
K.M. Rasch and  L. Mitas, Chem. Phys. Lett.,  \textbf{528}, 59 (2012); A. Kulahlioglu
and L. Mitas, submitted. 

\bibitem{qmcala}
J.J. Shepherd, A. Gruneis, G.H. Booth, G. Kresse, and A. Alavi, 
Phys. Rev. B \textbf{86}, 035111 (2012);
J.J. Shepherd, G. Booth, A. Gruneis, and A. Alavi, 
Phys. Rev. B  \textbf{85}, 081103 (2012); J.J. Shepherd, A. Alavi, private 
communication.

\bibitem{ecpbfd}
M. Burkatzki, C. Filippi, and M. Dolg,
J. Chem. Phys.  \textbf{126}, 234105 (2007).

\bibitem{g2new}
F. R. Petruzielo, J. Toulouse, C. J. Umrigar,
J. Chem. Phys. \textbf{136}, 124116 (2012).

\bibitem{chiesa}
S. Chiesa, D. M. Ceperley, R. M. Martin, and M. Holzmann,
Phys. Rev. Lett.  \textbf{97}, 076404 (2006).

\bibitem{feller}
D. Feller, K. A. Peterson, and J. G. Hill, J. Chem. Phys. \textbf{135}, 044102 (2011).

\bibitem{qmcmiso}
M. Bajdich, M. L. Tiago, R. Q. Hood, P. R. C. Kent, and F. A. Reboredo,
Phys. Rev. Lett.  \textbf{104}, 193001 (2010).

\bibitem{nda}
S. Hu, K. M. Rasch, and L. Mitas, in {\it Advances in Quantum Monte Carlo}, edited
by S. Tanaka, S. M. Rothstein, W. A. Lester, Jr., ACS Symposium Series, 
Vol. 1094, 2012, pp. 77.; arXiv:1307.5567.

%\bibitem{taut}
%M. Taut,
%Phys. Rev. A 
 %\textbf{48}, 3561 (1993)

%\bibitem{needs}
%N. Nemec, M. D. Towler, and R. J. Needs,
%J. Chem. Phys. \textbf{132}, 034111 (2010)

%\bibitem{arne}
%A. Luchow, et al, J. Chem. Phys. \textbf{XX}, (20XX).

\bibitem{jindrafeo}
J. Kolorenc and L. Mitas, Phys. Rev. Lett. \textbf{101}, 185502 (2008).



\end{thebibliography}
\end{document}